\documentclass{iopart}

\usepackage{latexsym,graphicx} 
\begin{document}

\title{Casimir forces between nano-structured particles}

\author{C E Rom\'{a}n-Vel\'{a}zquez and Bo E Sernelius}

\address{Dept. of Physics, Chemistry, and Biology, Link{\"o}ping
University, SE-581 83 Link{\"o}ping, Sweden}
\ead{roman@ifm.liu.se}

\begin{abstract}
We develop a computational study of Casimir forces between three
dimensional (3D) finite objects with an internal granular structure.  The
objects in the model consist of a finite arrangement of nanometer sized
spherical particles having a dipolar interaction.  In this model system one
can both study the basic properties of the Casimir forces, and the effects
from changing the parameters of the nano-structured materials that
constitute the particles; this last type of study leads to a form of
control of the Casimir force and an insight into possible technological
applications.  We present examples of both kinds of study.
\end{abstract}
In the last 20 years an increased interest in nano-systems has revealed
a new realm of physical phenomena.  These studies have gained special
attention due to the possibility of technological applications.  For
example, at nano-scale distances the dispersion forces play an important
role and one has considered the possibility to employ them in the development
of nano-mechanical devices.  Some first attempts have been
performed and several others have been proposed
\cite{nemscapasso,nems}.  Another special interest in nano-systems
comes from the fact that in nano-meter scale the physical properties of the
objects change with their size.  Nano particles are currently used as
building blocks of macroscopic materials that will have properties not
found in nature.  With the most modern techniques it has been possible to
produce arrangements of nano particles in 2D or 3D patterns with great
control of shape, size, and spacing.  With the control of these parameters
it has been feasible to accurately engineer the optical properties of these
so called metamaterials.  Some of the most amazing examples of new
artificial materials are those with negative refractive index.  They have
come in focus in the area of Casimir forces due to the possibility of
obtaining repulsive forces.  The use of this or other kind of Casimir
interactions for technological applications requires a deep understanding
of the basics of the phenomena and the experience of a wide range of ways
to control the interaction.

Here we consider a model of finite 3D objects consisting of ordered
arrangements of spherical particles with dipolar interaction.  This system
could be interpreted as a model of objects made of metamaterials.  We
present results of how the characteristics of the arrangement affects the
Casimir forces.  In particular we present examples of the use of
metamaterials to obtain Casimir rotatinal forces.  This model system can
also be used as a basic research tool to study the phenomenon of
Casimir forces due to a connection with a computational method called
discrete dipole approximation (DDA)\cite{DDA}.  In conditions where the
granularity of the material is not of relevance the results of the study of
this system could be applied to normal materials.  We present several
results that are consistent with well know results obtained from other
theoretical models.

The successful measurement of Casimir forces \cite{Lamoreaux} brought back
new attention to this effect predicted by the earlier quantum field theory. 
Some studies have been motivated by the possibility of technological
application in the form of nano-mechanical devices.  Different possible
interactions have been considered: lateral forces have already been
measured and rotational forces have been proposed in different
configurations \cite{nemscapasso,vdwnumeric}.  The Casimir forces show a
complex behaviour due to the material and surface geometry dependences.  An
important insight into the basics of the phenomena of Casimir forces has
been obtained from the various theoretical models \cite{revmilton}. 
However, these models rely on important simplifications that limit their
applicability to actual devices: perfect metal assumption, the $z$
translational invariance (2D problem) or non-retarded interactions. 
Although a computational model exists for the Casimir energy between
arbitrary 3D objects of real materials, due to the numerical complexity of
the problem, similar assumptions have been needed in order to produce
results \cite{Capassocomp}.  The unique results that exist at the moment
for these kind of systems, obtained with a general multipolar scattering
formalism \cite{Casphe}, are the interaction between two spheres and a
sphere above a substrate.  This kind of formalism gives results of high
accuracy for highly symmetric systems like pair of spheroids or cylinders
\cite{Casphe,mulgeometry,chargecyl}; for other geometries the
description becomes much more complex and the numerical calculations
slow and with low precision.  These problems hamper the use of these models
in a general study of how the Casimir forces depend on geometrical
shapes, configurations, materials, temperature, etc.  In what follows we
present a system that can be useful in the understanding and the
application of Casimir forces.

Consider a system of $N$ spherical particles, made of a homogeneous 
isotropic material characterized by the
dielectric function $\varepsilon\left(\omega\right)$, immersed in the 
vacuum. An electric
field $\mathbf{E}_{inc}(\bf {r},\omega)$ applied to the system induces a dipole moment
in each sphere as a consequence of its polarizability,
$\alpha\left(\omega\right)$,
\cite{Serneliusbook}\begin{equation}
\mathbf{p}_{j}(\omega)=\alpha\left(\omega\right)\left[\mathbf{E}_{inc}(\bf {r}_{j},\omega)-\sum_{k\neq
j}\mathbf{A}_{j,k}(\omega)\cdot\mathbf{p}_{k}(\omega)\right].\label{eq:
1}\end{equation}
The so called interaction tensor $\mathbf{A}_{j,k}$ gives the electric
field at the position $\mathbf{r}_{j}$ of particle $j$ due to the dipole
moment $\mathbf{p}_{k}$ of particle $k$ at position $\mathbf{r}_{k},$
\[
\mathbf{A}_{j,k}\left(\omega\right
)\cdot\mathbf{p}_{k}(\omega)=\frac{e^{iqr_{jk}}}{r_{jk}}\left\{
q^{2}\mathbf{r}_{jk}\times\left(\mathbf{r}_{jk}\times\mathbf{p}_{k}(\omega)\right)\right.+\]
\[
+\left.\frac{\left(1-iqr_{jk}\right)}{r_{jk}^{2}}\left[r_{ij}^{2}\mathbf{p}_{k}(\omega)-3\mathbf{r}_{jk}\left(\mathbf{r}_{jk}\cdot\mathbf{p}_{k}(\omega)\right)\right]\right\}
,\] where $ q=\omega/c$, with $ c $ the speed of light, and
  $\mathbf{r}_{jk}=\mathbf{r}_{j}-\mathbf{r}_{k}$, with $r_{jk}$ its
magnitude.  The polarizability of a sphere with radiative corrections
\cite{coradia} due to the retarded propagation of the electromagnetic
waves in its interior is given by
\[
\alpha\left(\omega\right)=a^{3}\frac{\varepsilon\left(\omega\right)-1}{3+\left[\varepsilon\left(\omega\right)-1\right]\left[1+\left(qa\right)^{2}-2i(qa)^{3}/3\right]}.
\]
The electromagnetic resonances are obtained as those frequencies $\omega_{s}$
that make the determinant of the linear system of equations for
$\mathbf{p}_k$, obtained from Eq. (\ref{eq: 1}), vanish, i.e., \[
0=G\left(\omega_{s}\right)=\det\left|\frac{1}{\alpha\left(\omega_{s}\right)}\delta_{\beta,\gamma}
\delta_{j,k}+\mathbf{A}_{j\beta,k\gamma}(\omega_{s})\right|,\]
where, $\beta,\gamma = x,y,z $, the coordinates of a cartesian system.
Using the argument theorem one obtains the formula for the zero point
energy \cite{Serneliusbook} \[
U\left(\omega\right)=-\frac{\hbar}{i2\pi}\int_{i0}^{i\infty}d\omega\log
G\left(\omega\right).\]
We calculate the interaction energy numerically by first considering
the whole system and then subtracting the result when the objects do not
interact.

\begin{figure}
\center
\includegraphics[width=8cm]{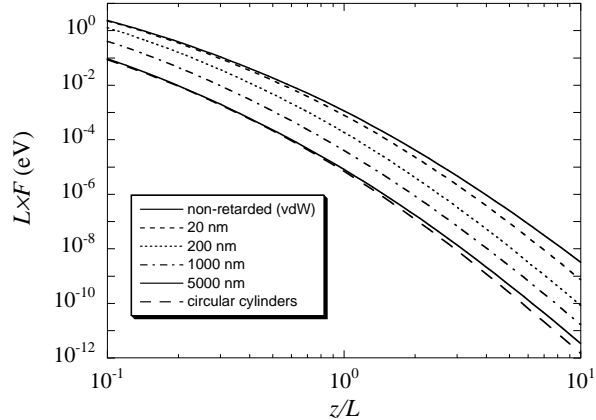}
\caption{Effect of the object size on the normal Casimir force.}

\end{figure}

In the figures that follow we present results for cubes and cylinders of
rectangular or circular cross sections.  All these objects are made up from
an ordered inner structure of small spheres.  In Fig. 1 we show results
of direct forces between two cubes of side length $L$.  We consider cubes
of different sizes but with the same number of gold spheres (1000) and
filling fraction $(4\pi/3^4)$.  We also present non-retarded van der Waals results ($c
\to \infty$ ) and of two interacting circular cylinders with their bases
aligned.  The area of the cylinder base is the same as for the cube of side
length $5\,\mu m$ but the height is $4\,\mu m$.  The experimental
dielectric function of gold has been taken from \cite{metaldata}.  In the
non-retarded calculations it is possible to choose one of the dimensions of
the system as a scale parameter due to the independence of the problem with
the dimensions of the system; we choose the side length of the cube, $L$. 
The retarded calculations are presented in the same way, but here one
obtains one curve for each cube size.  We observe that as expected the
bigger objects present bigger deviations from the vdW results due to the
bigger retardation effects within them, and also that the retardation
effects produce a weakening of the force even for the smallest cubes.  We
find for the bigger cubes at large distances the calculations including
retardation present the behavior of a retarded dipolar interaction $F\sim
z^{-8}$\cite{Serneliusbook}.  The non-retarded calculations show the known
power law behaviour of the form $F\sim z^{-7}$\cite{Serneliusbook}.  At
short distances we see that the smaller cubes present values similar to the
non-retarded calculations.  They vary as $F\sim z^{-3}$, consistent with a
pair of semi-infinite slabs
\cite{Serneliusbook}.  We refrain from doing comparisons with vdW results
at small distances for the bigger cubes because of the increasing effects
of granularity.  These effects become severe when the separation distance
is of the same order as the structural parameters of the cubes.  Finally,
we observe that, for the cylinder and the cube of $5\,\mu m$ the force is
very close.  This result is similar to that found for vdW forces in case of
homogeneous media where at shorter distances the force is independent of
the shape of the base of the cylinder
\cite{vdwnumeric}.

\begin{figure}
\center
\includegraphics[width=8cm]{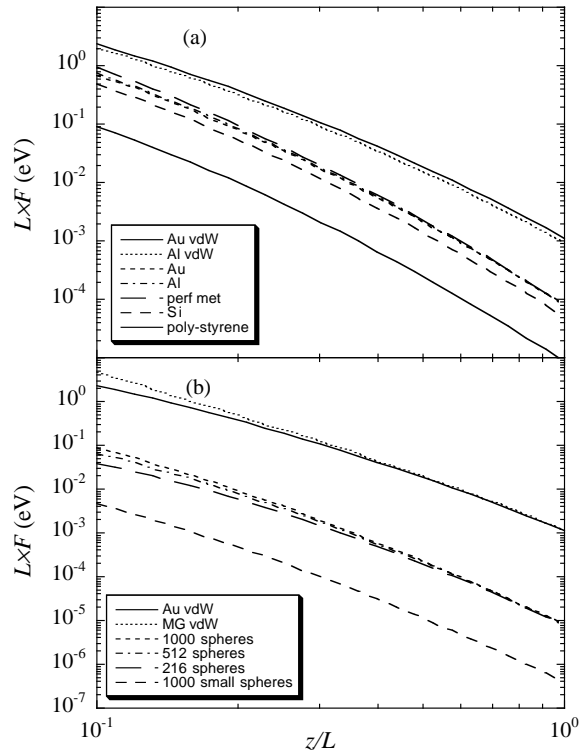}
\caption{Effects on the normal Casimir force between cubes from changes of nano-structure
parameters.  Retarded results are in the upper (lower) panel for $L = 0.5(5)\,
\mu m$); (a) different materials; (b) different geometric parameters}

\end{figure}

In Fig. 2 we explore the effects that the modification of the parameters of
the objects' nano-structure could have on the Casimir forces.  In panel (a)
we show the effects of the material of the spheres that constitute the
cubes and in panel (b) some of the geometry parameters of the arrangement. 
First we show the results for gold and aluminum \cite{metaldata} spheres
without considering retardation.  As can be seen the forces are different
at all distances.  However, in cubes of $0.5\,\mu m $ when retardation is
taken into account one obtains differences only at smaller distances.  At
larger distances both materials show the same force.  We also compare with
results from using the perfect metal approximation for the spheres. We observe
that with the perfect metal assumptions the forces show bigger differences
than those existing between real metals but eventually at larger distances
the differences go away.  We have observed that those differences are
smaller for bigger cubes.  Interestingly, the short distance forces do not
have the power law behavior corresponding to homogeneous perfect metal
objects, $F\sim z^{-4}$; instead, they show that of objects made of normal
dielectric materials.  Finally we show in the same panel results for
non-metallic materials \cite{edata}.  One observes differences at all
distances.  Smaller forces are obtained with silicon spheres and even much
smaller forces with poly-styrene spheres.

In panel (b) of Fig. 2, we first consider cubes of side length $5\,\mu m$
made with different number of spheres ($10^3, 8^3, 6^3$) but keeping the
same filling fraction $(4\pi/3^4)$.  We observe that at smaller distances,
when one expects stronger granularity effects, the forces in the three
different systems are different.  However, at larger distances their values
coincide.  This indicates that in this model at large distances the Casimir
force depends on the filling fraction of the material and not on the actual
size of the spheres.  A similar situation appears in the DDA model when the
convergence of the calculations is reached \cite{DDA}.  This fact is also
compatible with some of the results of the effective medium approximation
(EMA)\cite{EMA}.  In the EMA the objective is to determine the optical
properties of the system based on geometrical and material parameters of a
composite material.  Some of the simplest models of the EMA have the
filling fraction and the dielectric function of the materials as the only
information of their internal structure.  We show results from using the
formalism of Ref. \cite{vdwnumeric} to calculate the vdW force for
homogeneous objects with the dielectric function from an EMA formalism, the
Maxwell-Garnett model \cite{EMA}, based on similar assumptions as our
model. Note that they coincide in a wide range of
distances, and differ only at smaller distances where the effects of the
granularity of the model are stronger.  In Fig. 2(b) we also present
results for cubes of $10^3$ spheres with smaller radii ($L/50$).  A great
reduction of the force is observed.

\begin{figure}
\center
\includegraphics[width=8cm]{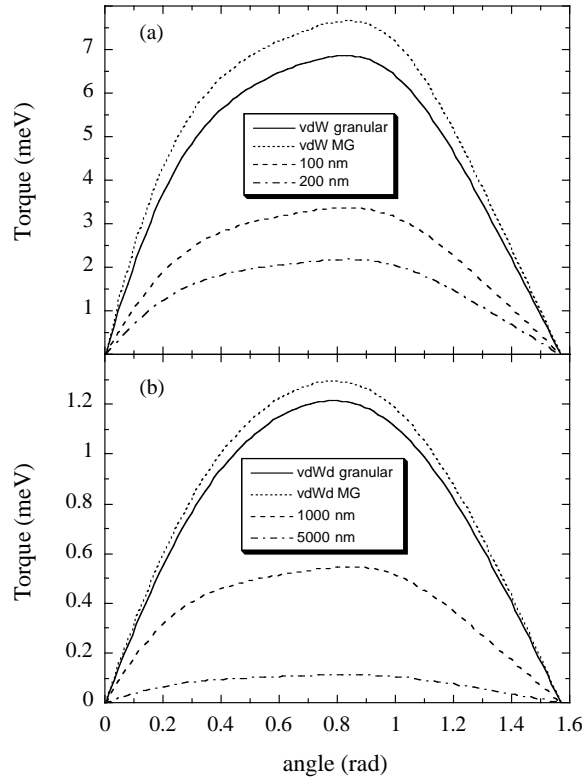}
\caption{Torque between rectangular cylinders: (a) $d_s=0.35L$; (b)
$d_s=0.35L$ in Casimir and $d_s=0.5L$ in vdW. }
\end{figure}

In Fig. 3 we show calculations of rotational forces between two finite
rectangular cylinders.  The lengths of the base are $L (1 \mu m)$ and $2L$ 
and the height is $0.5L$. 
They are kept at a distance $d_{s}$ and considered initially with their
bases aligned.  Then we consider a rotation around a perpendicular axis
that goes throughout their geometrical center. With a numerical derivation
we calculate the torque on the system.  We consider results for objects of
different lengths $L$ and also different separations $d_{s}$ and compare
with the homogeneous model of \cite{vdwnumeric} using the
Maxwell-Garnett dielectric function model.  Note that the homogeneous and
discrete model present similar shapes, despite of the different approaches. 
As we see in panel (b) for the bigger value of $d_s$ the numerical differences 
of both non-retarded
calculations are smaller; this is expected from the
experience of normal forces at greater separation distances.  These last results
are another evidence for that the model produces results that are
compatible with those obtained from accurate models of homogeneous objects.
\begin{figure}
\center
\includegraphics[width=8cm]{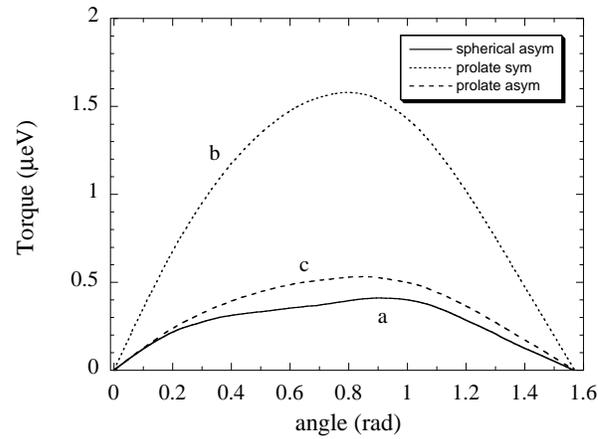}
\caption{Torque between circular cylinders of anisotropic 
metamaterials.(a) spheres in asymmetric lattice;
(b) prolates in symmetric lattice; (c) prolates in asymmetric lattice.}
\end{figure}
In the following we present an example of Casimir rotational forces between
two cylinders made of anisotropic metamaterials.  The effect of rotational
interation between two bi-refringent material slabs has been studied in
\cite{nemscapasso}.  In this phenomenon it is interesting to, instead of
being limited by the availability of natural birefringent materials,
consider metamaterials as a way to increase or control these forces.  In
Fig. 4 we show results for different kinds of anisotropic metamaterials
obtained from our model by changing the parameters of the constituents. 
Two anisotropic arrangements of particles produce a difference of energy
when their axes are not aligned; and as a consequence a rotation.  A pair
of finite cylinders of height $ 0.43\, \mu m$ are positioned so that their
circular bases of area $A (1\,\mu m^2)$ are kept parallel at a distance of
$0.4A^{1/2}$.  We consider three different anisotropic inner structures:
one with spheres of radii $A^{1/2}/42$ in a lattice obtained from a cubic
one by increasing one of its lattice parameters ($A^{1/2}/14$) by a factor
of $1.2$; the second arrangement considers prolate spheroids with aspect
ratio $1.2$ and the bigger semi-axis of $A^{1/2}/42$ in a cubic lattice;
the third is a combination of both with the long axis of the spheroid in
the direction of the modified lattice direction.  This last case resembles
the production of prolate nanoparticles by stretching a sample of metallic
spheres immersed in a dielectric matrix.  The prolate polarizability was
taken from \cite{mulgeometry}.  We see that the maximum torque is obtained
with the prolates in a cubic lattice, while the smaller is obtained with
the spheres in the stretched lattice.  However the combinations of both
geometries do not show a greater anisotropy.  The effect of the stretched
lattice compensates the effect of the prolate spheroids.

In summary, we have introduced a simple model system of two nano-structured
3D objects for the study of dispersion forces with the inclusion of
retardation and material effects.  This system could be interpreted as a
model of objects made of metamaterials, in which it is possible to control
the magnitude and direction of Casimir forces through the modification of
the internal structure.  Furthermore, it can be used to study the effects
of shape, size, temperature and materials on the Casimir forces between
compact 3D objects; due to the complexity of the theoretical models
available, this has not been possible until now.  This letter is just a
very brief summary of our results.  A more detailed publication will
follow.

This research was sponsored by EU within the EC-contract No:012142-NANOCASE
and support from the VR Linn\'{e} Centre LiLi-NFM and from CTS is
gratefully acknowledged.


\begin{thebibliography}{10}
\bibitem{nemscapasso} V. A. Parsegian and G. H. Weiss, \textit{J.
Adhe. }\textbf{3}, 259 (1972); F. Capasso, J. N. Munday, D. Iannuzzi
and H. B. Chan, \textit{IEEE Sel. T. Quant. Elect. }\textbf{13}, 400
(2007).

\bibitem{nems} C. H. Ke, N. Pugno, B. Peng, H. D. Espinosa, \textit{J.
Mech. Phys. Solids }\textbf{53,} 1314 (2005); A. Ashourvan,
M. F. Miri and R. Golestanian, \textit{Phy. Rev. Lett.} \textbf{98,} 
140801 (2007).

\bibitem{DDA} B. T. Draine, \textit{Astrophys. J.}\textbf{333}, 848 (1988).

\bibitem{Lamoreaux} S. K. Lamoreaux, \textit{Phys.
Rev. Lett. }\textbf{78}, 5 (1997).

\bibitem{vdwnumeric} C. E. Rom\'{a}n-Vel\'{a}zquez and B. E. Sernelius, \textit{J. Phys. A}
\textbf{41}, 164008 (2008).

\bibitem{revmilton}see, e.g. , K. A. Milton, \textit{J. Phys. A. }\textbf{37},
R209 (2004).

\bibitem{Capassocomp}A. Rodriguez, M. Ibanescu, D. Iannuzzi, F. Capasso,
J. D. Joannopoulos, and S. G. Johnson, \textit{Phy. Rev. Lett.} \textbf{99},
080401 (2007).

\bibitem{Casphe}T. Emig, N. Graham, R. L. Jaffe and M. Kardar \textit{Phy.
Rev. Lett.} \textbf{99}, 170403 (2007).

\bibitem{mulgeometry}C. Noguez and C. E. Rom\'{a}n-Vel\'{a}zquez,
\textit{Phys. Rev. B} \textbf{70}, 195412 (2004);
C. E. Rom\'{a}n-Vel\'{a}zquez and C. Noguez, \textit{J. Phys. A} \textbf{39},
6695 (2006).

\bibitem{chargecyl}T. Emig, R. L. Jaffe, M. Kardar and A. Scardicchio,
\textit{Phys. Rev. Lett.} \textbf{96}, 080403 (2006);
F. D. Mazzitelli, D. A. R. Dalvit and F. C. Lombardo,
\textit{New J. Phys.} \textbf{8}, 240 (2006).

\bibitem{Serneliusbook} see, e.g., B. E. Sernelius\textit{, Surface Modes in
Physics} (Wiley-VCH, Berlin, 2001).

\bibitem{coradia}M. Meier and A. Wokaund, \textit{Opt. Lett. }\textbf{8},
581 (1983).

\bibitem{metaldata} A. Lambrecht and S. Reynaud, \textit{Eur. Phys. 
J. D} \textbf{8},
309 (2000).

\bibitem{edata}L. Bergstr\"{o}m, \textit{Adv. Colloid Interfac. }
\textit{70}, 125 (1997).

\bibitem{EMA}see, e.g., G. A. Niklasson, C. G. Granqvist and O. 
Hunderi, \textit{Appl.
Opt.} \textbf{20}, 26 (1981).
\end{thebibliography}
\end{document}